\documentclass{osa-article}

\journal{osac}
\usepackage[normalem]{ulem}
\usepackage[utf8x]{inputenc} 
\usepackage{graphics}

\begin{document}

\title{Fully phase-stabilized 1 GHz turnkey frequency comb at 1.56~{\textmu}m}

\author{Daniel M. B. Lesko,\authormark{1,2,$\dagger$,*} Alexander J. Lind,\authormark{1,3,$\dagger$} Nazanin  Hoghooghi,\authormark{4,$\dagger$} Abijith Kowligy,\authormark{1,3} Henry Timmers,\authormark{1} Pooja Sekhar,\authormark{1,3}  Benjamin Rudin,\authormark{5} Florian Emaury,\authormark{5} Gregory B. Rieker,\authormark{4} and Scott A. Diddams\authormark{1,3,*}}

\address{\authormark{1}Time and Frequency Division, NIST, 325 Broadway, Boulder, Colorado 80305, USA\\
\authormark{2}Department of Chemistry, University of Colorado, 215 UCB, Boulder, Colorado 80309, USA\\
\authormark{3}Department of Physics, University of Colorado, 2000 Colorado Ave., Boulder, Colorado 80309, USA\\
\authormark{4}Precision Laser Diagnostics Laboratory, Department of Mechanical Engineering, University of Colorado
Boulder, Boulder, CO 80309, USA\\
\authormark{5}Menhir Photonics AG, Thiersteinerallee 71, CH-4053 Basel, Switzerland\\
\authormark{$\dagger$}These authors contributed equally to this work}

\email{\authormark{*}Daniel.Lesko@nist.gov, Scott.Diddams@nist.gov}


\begin{abstract}
Low noise and high repetition rate optical frequency combs are desirable for many applications from timekeeping to precision spectroscopy. 
For example, gigahertz repetition rate sources greatly increase the acquisition speed of spectra in a dual-comb modality when compared to lower repetition rate sources, while still maintaining sufficient instantaneous resolution to resolve ro-vibrational signatures from molecules in a variety of conditions. 
In this paper, we present the stabilization and characterization of a turnkey commercial 1~GHz mode-locked laser that operates at telecom wavelengths (1.56~{\textmu}m). Fiber amplification and spectral broadening result in the high signal-to-noise ratio detection and stabilization of \textit{f}$_{\textit{ceo}}$ with 438~mrad of residual phase noise (integrated from 10$^2$~to~10$^7$~Hz). Simultaneously, we stabilize the beatnote between the nearest comb mode and a cavity stabilized continuous-wave laser at 1.55~{\textmu}m with 41~mrad of residual phase noise (integrated from 10$^2$~to~10$^7$~Hz). This robust, self-referenced comb system is built with off-the-shelf polarization-maintaining fiber components and will be useful for a wide range of low noise frequency comb applications that benefit from the increased repetition rate.
\end{abstract}

\section{Introduction}

Optical frequency combs (OFCs) provide a phase-coherent connection between oscillators across the full electromagnetic spectrum, from radio to optical frequencies\cite{hall_nobel_2006,hansch_nobel_2006}. This has enabled unique applications such as the readout and intercomparison of the next generation of optical atomic clocks\cite{diddams_optical_2001,rosenband_frequency_2008}, exceedingly low-noise microwave generation\cite{fortier_generation_2011,xie_photonic_2017,giunta_compact_2020,nakamura_coherent_2020}, precise free-space time transfer\cite{giorgetta_optical_2013}, and accurate broadband spectroscopy \cite{thorpe_broadband_2006,coddington_coherent_2008,kowligy_infrared_2019,metcalf_stellar_2019}.

Frequency comb spectroscopy is often performed in a dual-comb configuration which necessitates low phase noise performance\cite{coddington_dual-comb_2016}. Dual-comb spectroscopy (DCS) has been used in an assortment of applications including
broadband long-distance standoff detection\cite{rieker_frequency-comb-based_2014}, as well as cavity-enhanced \cite{fleisher_coherent_2016-2} and nonlinear spectroscopy\cite{ideguchi_coherent_2013}, to name a few.  
In DCS, there is a trade off between the instantaneous resolution given by the comb mode spacing, and spectral acquisition speed, which scales with the square of the repetition rate\cite{coddington_dual-comb_2016}.  Depending on the spectroscopic sample that is being studied, frequency combs with mode spacing from 10~MHz to 100~GHz may be appropriate, and a variety of frequency comb platforms are being developed along these lines.  One interesting regime is gas-phase spectroscopy at high-temperatures and pressures (such as combustion), where a mode spacing of 1~GHz provides sufficient spectral sampling of ro-vibrational lines\cite{hoghooghi_11-mus_2020}, while at the same time enabling acquisition over spectral bandwidths of 10~THz at rates approaching 100~kHz.  Beyond these focused benefits for spectroscopy, frequency combs at 1 GHz repetition rate fall in a convenient operational space for other metrology applications by providing increased power per comb tooth (compared to 100~MHz), while still maintaining simplicity when it comes to electronic interfaces, digital sampling, and nonlinear spectral broadening in optical fiber (compared to 10~GHz and higher). 

Over the past two decades there has been significant work on self-referenced frequency combs sources with $\sim$1 GHz mode spacing.  These include combs built around laser gain provided by titanium at 0.8~{\textmu}m \cite{diddams_compact_2001,fortier_octave-spanning_2006,chen_octave-spanning_2008}, ytterbium at 1~{\textmu}m \cite{hartl_fully_2009, pekarek_self-referenceable_2011, endo_direct_2015,hakobyan_carrier-envelope_2017,hakobyan_full_2017}, erbium at 1.56~{\textmu}m \cite{chao_self-referenced_2012,shoji_ultra-low-noise_2016}, and chromium at 2.35~{\textmu}m\cite{smolski_half-watt_2018}.  Of all these options, 1.56~{\textmu}m is unique because technical developments driven by the telecommunications industry have resulted in inexpensive pump diodes, a wide range of off-the-shelf fiber components, dispersion compensating fibers, and commercial highly nonlinear fibers for spectral broadening. At lower repetition rates, this has enabled robust erbium fiber based frequency combs that can be used outside the lab\cite{sinclair_operation_2014,lezius_space-borne_2016} as well as for low noise long term comb operation for frequency measurements\cite{inaba_long-term_2006,schibli_frequency_2004}.  A noteworthy advancement at the 1~GHz repetition rate is a design built around a monolithic CaF cavity and 1.56~{\textmu}m erbium-glass gain medium that resulted in ultra-low phase noise\cite{shoji_ultra-low-noise_2016}.

Building on these advances and  motivations, in this paper, we demonstrate a fully phase stabilized frequency comb that utilizes a commercial turnkey 1~GHz mode-locked laser operating at 1.56~{\textmu}m. Starting with the fiberized laser output, we employ a polarization maintaining (PM) dispersion managed amplifier\cite{fermann_self-similar_2000} and PM highly nonlinear fiber (HNLF) to broaden the spectrum to an octave that spans from 1~{\textmu}m to 2~{\textmu}m. With this robust octave spanning source, we measure the carrier envelope offset frequency (\textit{f}$_{\textit{ceo}}$) via an inline f-to-2f interferometer with nearly 40~dB signal to noise ratio (SNR) at 300~kHz resolution bandwidth (RBW) and phase lock it to an RF reference with 438~mrad of integrated phase noise (10$^2$~to~10$^7$~Hz). We then phase lock one tooth of the comb to a cavity stabilized laser at 1.55~{\textmu}m resulting in an integrated phase noise of 40.6~mrad (10$^2$~to~10$^7$~Hz). This fully phase stabilized robust and turnkey source should be useful for applications including timing distribution, ultra-low noise microwave generation, optical clock comparisons, and fast-acquisition spectroscopy.

\section{Experimental Setup and Results}

\begin{figure}[ht!]
\centering\includegraphics[width=0.7\textwidth]{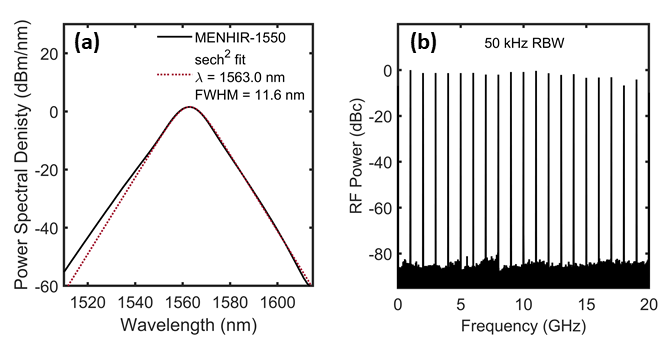}
\caption{\textbf{(a)}, Spectrum from the 1 GHz laser oscillator (black) and sech$^{2}$ fit (red dashed) showing an 11.6~nm full width at half maximum (FWHM) at 1563.0~nm. \textbf{(b)}, RF spectrum of the output of a fast photodiode, showing clean harmonics of the fundamental 1 GHz repetition rate.
}
\label{fig:osc}
\end{figure}

The mode locked laser we use for this work is a commercially-available oscillator at 1.56~{\textmu}m (MENHIR-1550; hereinafter referred to as OFC when stabilized or simply 1~GHz laser). The turnkey laser is in a sealed housing with output provided though an integrated PM fiber. It is powered with 5~V DC and draws <2~A current. The oscillator is passively mode locked, providing a sech$^2$ optical spectrum with pulses at 1~GHz, shown in Fig.~\ref{fig:osc}.  The duration of the chirped pulses at the end of the output fiber is approximately 300~fs. 
For the experiments that follow, we essentially treat this mode locked laser as a ``black box" from which a fully-stabilized 1 GHz frequency comb is created. 

\begin{figure}[ht!]
    \centering\includegraphics[width=0.7\textwidth]{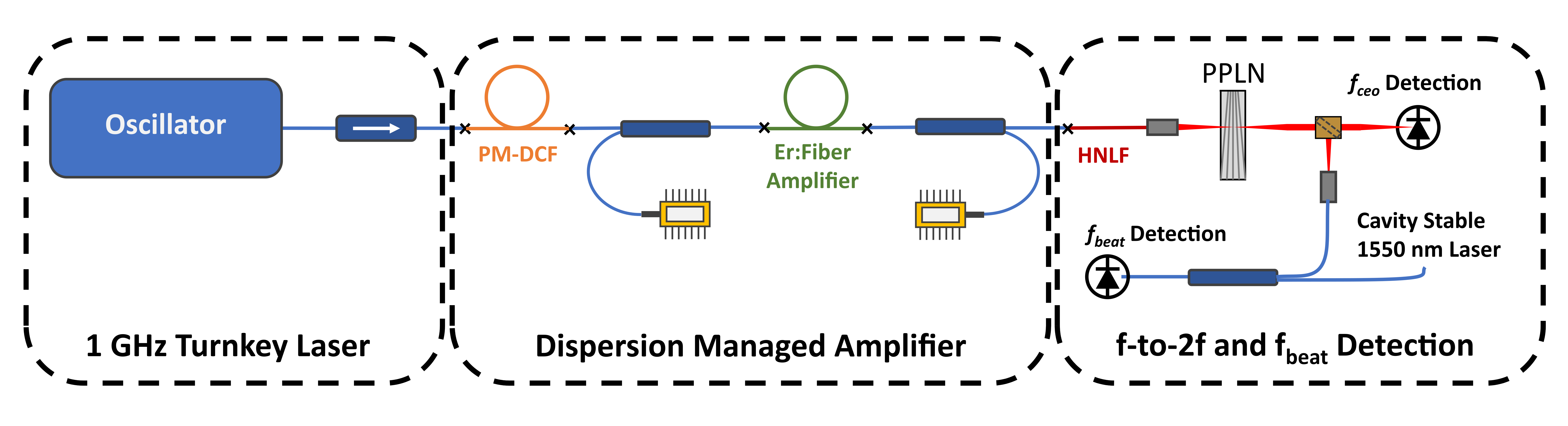}
\caption{Experimental setup used for \textit{f}$_{\textit{beat}}$ and \textit{f}$_{\textit{ceo}}$ detection. After the turnkey oscillator, the pulses are amplified in a dispersion managed amplifier and spectrally broadened for f-to-2f measurement in periodically poled LiNbO$_3$ (PPLN). Both the 1~{\textmu}m and 1.55~{\textmu}m components are spectrally separated to allow for simultaneous locking of \textit{f}$_{\textit{ceo}}$ and \textit{f}$_{\textit{beat}}$.}
\label{fig:outline}
\end{figure}

The average output power of the oscillator is 60~mW, providing sufficient power to seed multiple optical amplifiers.
To fully phase stabilize the 1 GHz laser we start by building a dispersion managed short pulse amplifier, and an inline f-to-2f interferometer. This design, Fig.~\ref{fig:outline}, allows us to measure \textit{f}$_{\textit{ceo}}$ as well as simultaneously beat the comb against a cavity stabilized CW laser for optical phase locking. The oscillator output, amplifier, and broadening stages all employ PM fiber and are designed with commercial components, allowing for reproducible and robust day-to-day operations.

The oscillator's output was sent through an isolator and 50/50 coupler (for the use of the comb in other experiments) and then into a dispersion-managed amplifier. 
With an input power of 12~mW, the gain of the amplifier was 17.5~dB providing an output power of 680~mW.
Dispersion management before the amplifier was done with $\sim 1.5$~m of PM dispersion compensation fiber, Thorlabs PM-DCF with dispersion D~=~-100~$\text{ps}/(\text{nm}\cdot \text{km})$, and $\sim 3$~m of PM-1550 fiber.
This amplifier was forward and backward pumped by four 1~W, 980~nm single mode pump diodes that are polarization multiplexed in pairs. We used 2~m of LIEKKI Er80-4/125-HD-PM gain fiber and the least amount of PM-1550 possible on the wavelength-division multiplexers.
The input pulse chirp was optimized by adjusting the PM-1550 fiber length for a slight anomalous chirp into the amplifier to allow temporal compression and  spectral broadening by self phase modulation (SPM) in the  gain fiber of the amplifier.

\begin{figure}[ht!]
\centering\includegraphics[width=0.7\textwidth]{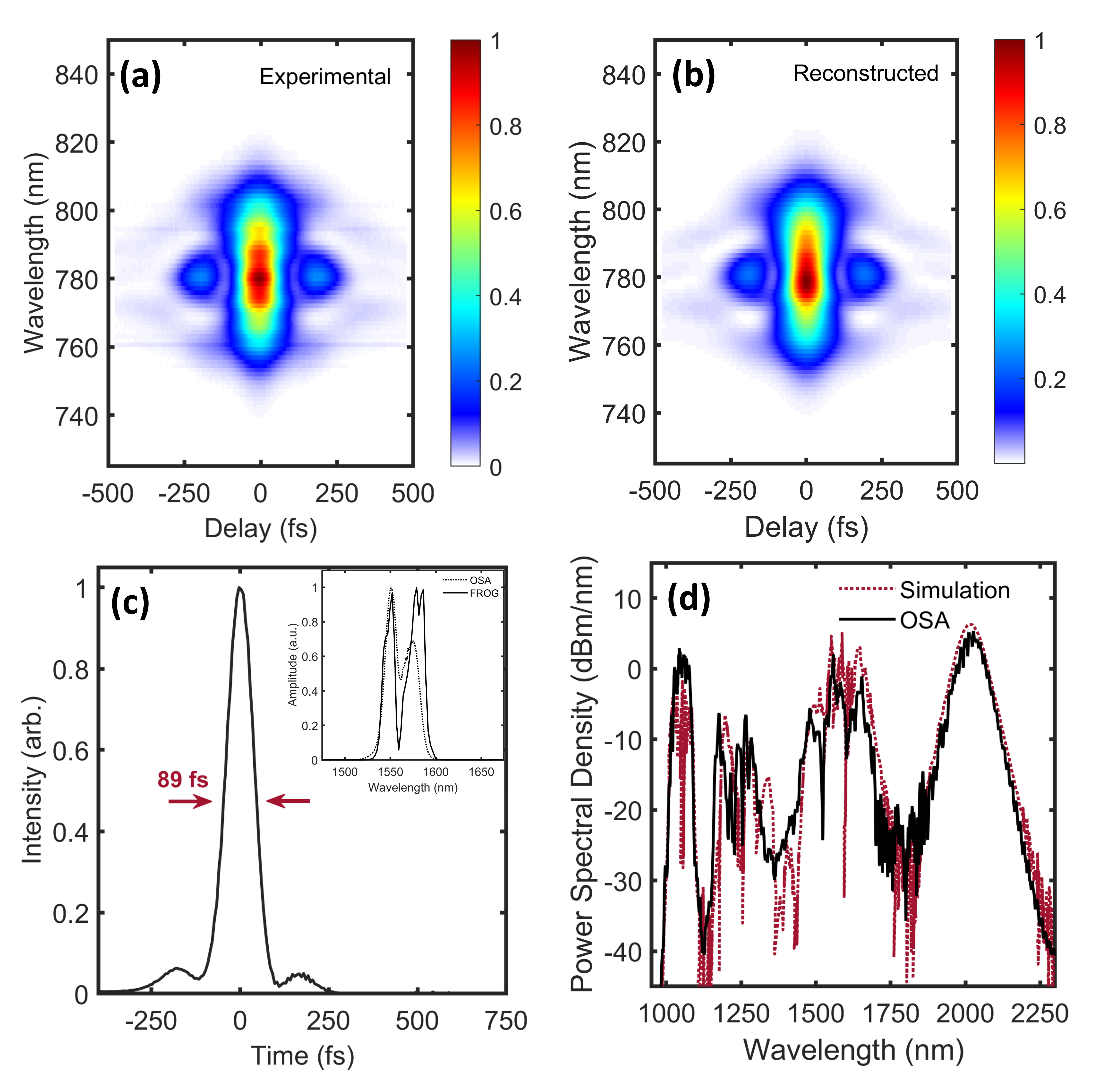}
\caption{\textbf{(a)}, Experimental second harmonic generation frequency resolved optical gating (SHG-FROG) measured at the output of the dispersion managed amplifier. \textbf{(b)}, Reconstructed SHG-FROG with an error of 0.5\%. \textbf{(c)}, Temporal profile of the reconstructed pulse (inset: spectra from an optical spectrum analyzer (OSA) and FROG reconstruction). \textbf{(d)}, Corresponding octave spanning spectrum and simulation results from the nonlinear Schr\"odinger equation (NLSE) after broadening in HNLF.}
\label{fig:amp}
\end{figure}

The amplified pulse, which is normally chirped by the gain fiber, was compressed in 40~cm of PM-1550 fiber and its duration was measured via second harmonic generation frequency resolved optical gating (SHG-FROG). The retrieved pulse is presented in Fig.~\ref{fig:amp} a-c and has a width of 89~fs, which then drives supercontinuum in the HNLF. For a strong \textit{f}$_{\textit{ceo}}$ beat, a high power spectral density component is needed at 2~{\textmu}m for generating second harmonic to overlap with the dispersive wave at 1~{\textmu}m.
We simulate propagation with the nonlinear Schr\"odinger equation (NLSE) and settle on 1.7~m of HNLF, D~=~5.7~$\text{ps}/(\text{nm}\cdot \text{km})$, to achieve the spectrum of Fig.~\ref{fig:amp} d, which includes a soliton component shifted to 2~{\textmu}m for self referencing.
The output of the amplifier is focused into a 1~mm periodically poled lithium niobate (PPLN, with poling period $\Lambda$= 30.80~{\textmu}m) to provide second harmonic generation at 2~{\textmu}m. 
After the PPLN, the short wave component ($\leq$1.2~{\textmu}m) is spectrally separated and used for \textit{f}$_{\textit{ceo}}$ detection.
The remaining light is used for optically phase locking \textit{f}$_{\textit{rep}}$.

\begin{figure}[ht!]
\centering\includegraphics[width=0.7\textwidth]{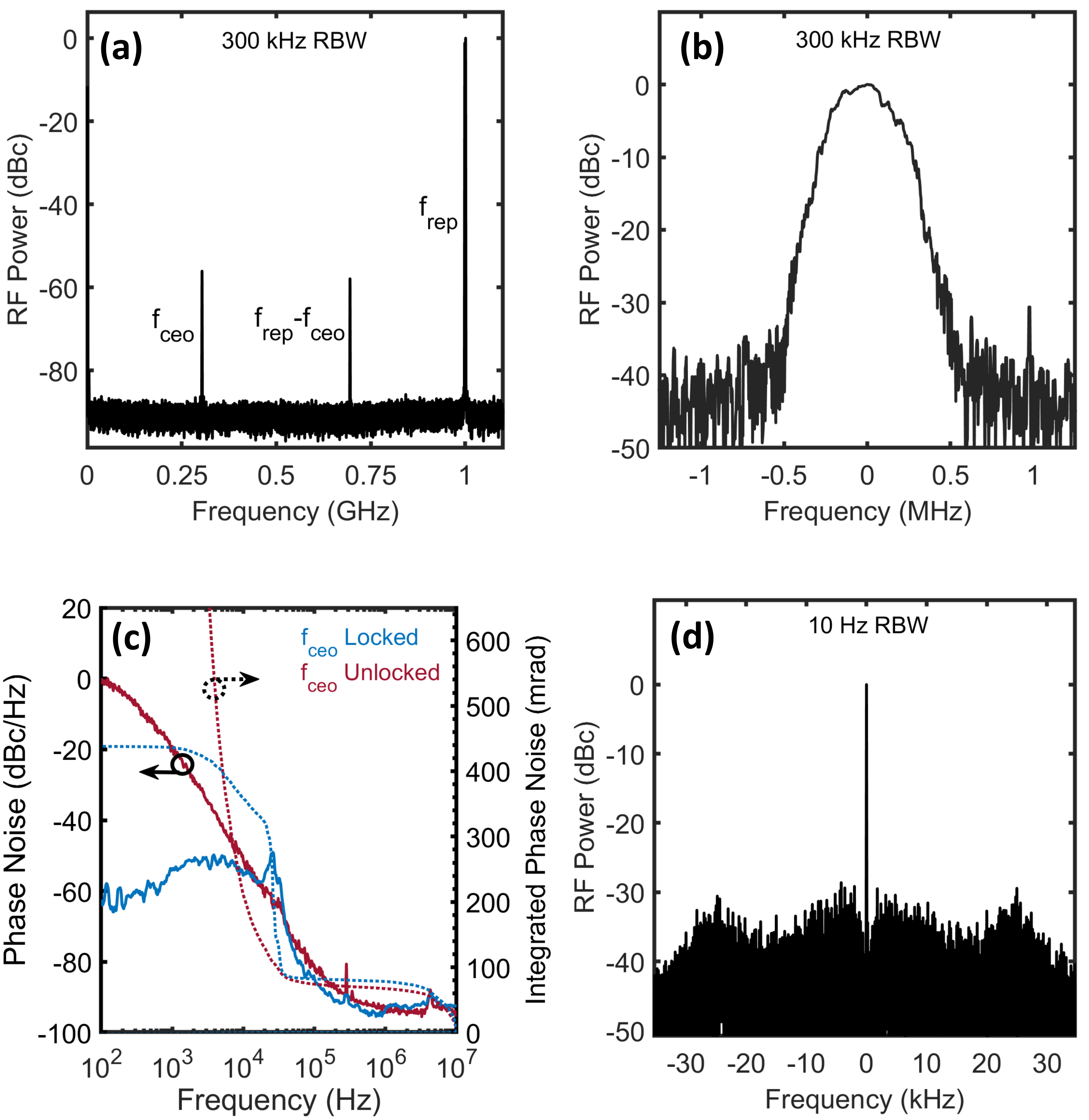}
\caption{\textbf{(a)}, RF spectrum from DC to 1.2~GHz showing \textit{f}$_{\textit{ceo}}$, \textit{f}$_{\textit{rep}}$-\textit{f}$_{\textit{ceo}}$, and \textit{f}$_{\textit{rep}}$ at 300 kHz RBW. \textbf{(b)}, RF spectrum of \textit{f}$_{\textit{ceo}}$ showing 40~dB SNR at 300~kHz RBW. \textbf{(c)}, Solid lines: \textit{f}$_{\textit{ceo}}$ phase noise power spectral density, dotted lines: integrated phase noise from high Fouier frequencies to DC (total: 438 mrad integrated from 10$^2$ to 10$^7$~Hz). \textbf{(d)}, RF spectrum of \textit{f}$_{\textit{ceo}}$ locked with 10~Hz RBW.}
\label{fig:fceo}
\end{figure}

The shortwave components were focused onto a fast photodiode, and the corresponding RF spectrum (DC~-~1.2~GHz) is shown in Fig~\ref{fig:fceo}~a.
The photodiode output was bandpass filtered at \textit{f}$_{\textit{rep}}$-\textit{f}$_{\textit{ceo}}\approx 700$ MHz, and amplified to provide 40~dB SNR at 300~kHz RBW, shown in Fig.~\ref{fig:fceo}~b. 
This high SNR shows that commercial and robust fiber technology can be easily adapted for measurement \textit{f}$_{\textit{ceo}}$ on the 1~GHz oscillator.
The \textit{f}$_{\textit{ceo}}$ signal is then electronically divided by 32 and phase locked to a signal at 22~MHz from a low phase noise RF generator referenced to a hydrogen maser. The phase lock was completed with feedback to the pump current of the 1~GHz laser.  The sensitivity of this tuning of \textit{f}$_{\textit{ceo}}$ was 0.78~$\frac{\text{MHz}}{\text{V}}$, and a larger tuning range of 80~MHz could be achieved by changing the DC pump current, while still maintaining a mode-locked state. When locked, the residual phase noise on \textit{f}$_{\textit{ceo}}$ was 438~mrad, integrated across 10$^2$-10$^7$~Hz (Fig.~\ref{fig:fceo}~c).
The pump current modulation bandwidth is 20~kHz, as shown by the servo bump present in the phase noise power spectral density (Fig.~\ref{fig:fceo}~c).
The locked \textit{f}$_{\textit{ceo}}$ is shown in Fig.~\ref{fig:fceo}~d with 70~kHz span and a 10~Hz RBW. 
The \textit{f}$_{\textit{ceo}}$ of a second 1~GHz laser was measured using the same amplifier and f-to-2f interferometer. Both oscillators had nearly identical spectra, pulse duration and intracavity dispersion, resulting in similarly high SNR \textit{f}$_{\textit{ceo}}$ signals for the same dispersion managed amplifier and f-to-2f interferometer.

With a tight lock on \textit{f}$_{\textit{ceo}}$, we simultaneously optically phase lock one mode of the 1~GHz laser to a CW frequency-stabilized laser at 1550~nm. This effectively controls \textit{f}$_{\textit{rep}}$ with high sensitivity since the frequency of a comb mode near 1550~nm is  n$\cdot$\textit{f}$_{\textit{rep}}$, where n is on order 2$\times$10$^5$. To generate the beat with the CW laser, we couple the residual spectrum around 1550~nm back into fiber, and measure a heterodyne beat with the CW light on a fast photodiode. In similar fashion to the stabilization of \textit{f}$_{\textit{ceo}}$, the heterodyne beat \textit{f}$_{\textit{beat}}$ is also phase locked to a signal from a low phase noise RF generator. However, in this case no additional frequency division was required. A fast piezo transducer on a cavity element is used to control the repetition rate. In fact, the 1~GHz laser has two intracavity piezos, allowing for fast/fine and slow/coarse modulation of the repetition rate over a range of 200~kHz with a modulation bandwidth close to 100~kHz.
The fast and coarse piezos have a sensitivity of 3.72~$\frac{\text{Hz}}{\text{V}}$ and 1541.4~$\frac{\text{Hz}}{\text{V}}$ respectively.

\begin{figure}[ht!]
\centering\includegraphics[width=0.7\textwidth]{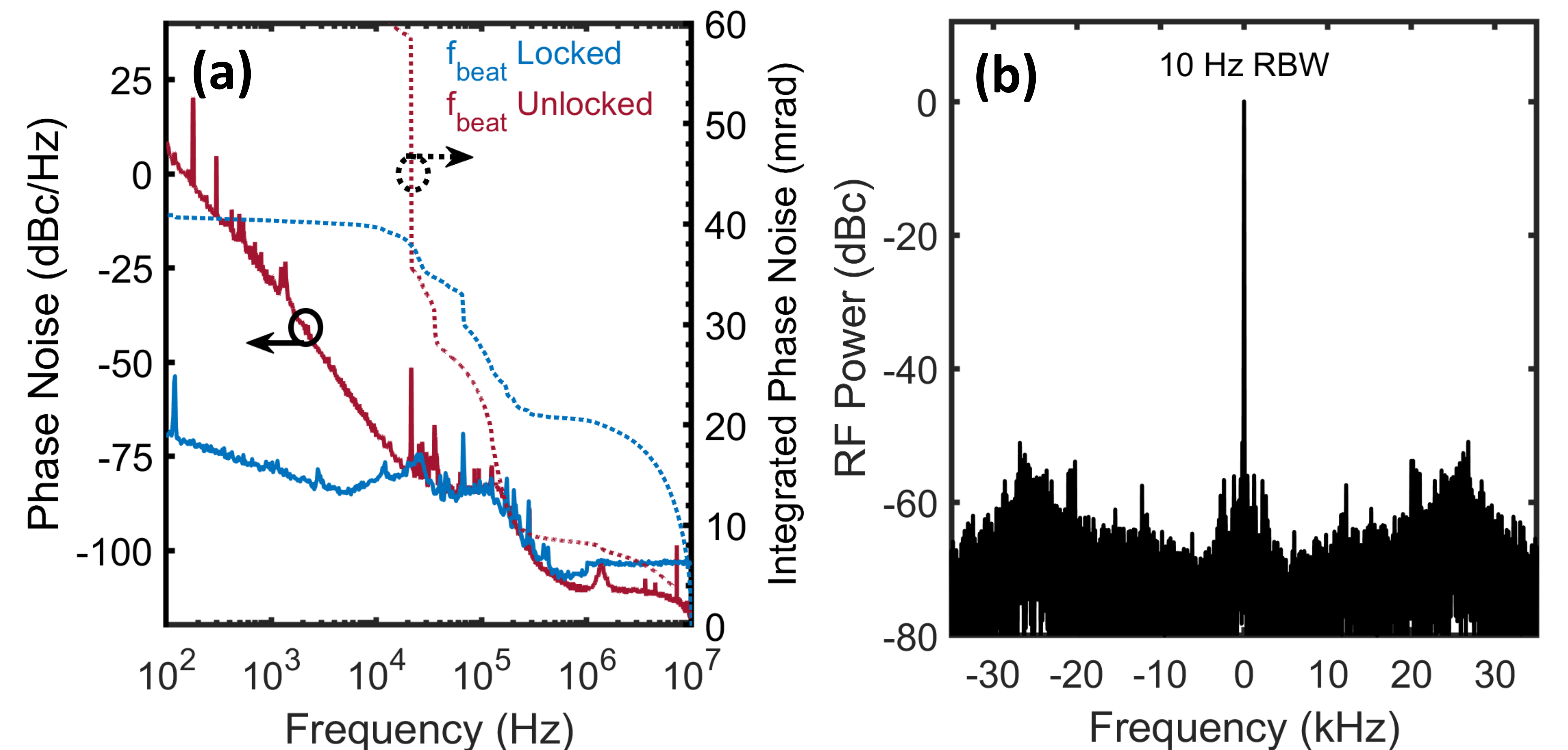}
\caption{\textbf{(a)}, Solid lines: \textit{f}$_{\textit{beat}}$ phase noise power spectral density, dotted lines: integrated phase noise from high Fourier frequencies to DC (total: 40.6~mrad integrated from 10$^2$~to~10$^7$~Hz). \textbf{(b)}, RF spectrum of \textit{f}$_{\textit{beat}}$ locked at 10~Hz RBW.}
\label{fig:fbeat}
\end{figure}

With the servo loop closed, the residual phase noise on \textit{f}$_{\textit{beat}}$ is 40.6~mrad, integrated across 10$^2$~to~10$^7$~Hz, as shown in Fig.~\ref{fig:fbeat} a. 
The peak at 20~kHz in the phase noise power spectral density is due to cross-coupling from the servo locking \textit{f}$_{\textit{ceo}}$.
The optically locked beat is shown in Fig.~\ref{fig:fbeat} b with a 70~kHz span and a 10~Hz RBW.
While the center frequency of the 1550~nm CW laser is stable at the 1 Hz level, it has excess frequency noise >30~kHz, such that the plateau in the noise spectrum of Fig.~\ref{fig:fbeat} near 100~kHz and the peak near 1.2~MHz are not due to the OFC. 
This implies that the limit on the integrated phase noise \textit{f}$_{\textit{beat}}$ could be less than 40.6~mrad.

\section{Conclusion}

In summary, we fully phase stabilized a robust, turnkey 1~GHz oscillator.
To do this, we built a PM dispersion managed amplifier with 17.5~dB of gain yielding 89~fs pules with 7.4~kW of peak power. 
With these short pulses, we are able to efficiently broaden in highly nonlinear fiber to achieve an octave of spectral bandwidth in order to stabilize the carrier envelope offset frequency.
By careful fiber length management, an inline f-to-2f interferometer gives an \textit{f}$_{\textit{ceo}}$ beat with 40~dB SNR at 300~kHz RBW, which is phase locked to a low noise RF source with 438~mrad of phase noise integrated from 10$^2$~to~10$^7$~Hz. 
The residual 1550 nm light from the f-to-2f interferometer is used to phase lock one tooth of the comb to a cavity stabilized laser with 40.6 mrad of phase noise integrated from 10$^2$~to~10$^7$~Hz.

In recent years, 1550~nm telecom OFCs have proved to be useful low noise pumps for nonlinear frequency conversion to the mid infrared (3~to~25~{\textmu}m) frequency combs for applications such as spectroscopy in both the fingerprint (5~to~25~{\textmu}m)\cite{timmers_molecular_2018,kowligy_infrared_2019,sell_phase-locked_2008,lee_milliwatt_2020} and functional group (3~to~5~{\textmu}m)\cite{lind_mid-infrared_2020,ycas_high-coherence_2018,sell_field-resolved_2008} regions. These sources can provide non-destructive high SNR measurements on chemical composition and reaction mechanisms for various processes\cite{griffiths_fourier_2006}. This turnkey OFC higher repetition rate would enable 100 times faster data acquisition rates than a 100~MHz source when operated in a dual-comb configuration\cite{coddington_dual-comb_2016}, while still providing access to spectroscopic absorption features relevant to combustion and atmospheric chemistry\cite{draper_broadband_2019}. 
We see a clear path for utilizing these turnkey 1~GHz OFCs for performing mid-infrared spectroscopy on combustion systems.

\section*{Acknowledgments}
Commercial items and parts are identified in this paper for informational purposes only. Such identification does not imply recommendation or endorsement by the National Institute of Standards and Technology, nor is it intended to imply that the products identified are necessarily the best available for the purpose.
The authors thank Thomas Schibli for his contributions, as well as Sida Xing and Nima Nader for their helpful comments. DMBL and AK are supported by the University of Colorado through award 70NANB18H006 from the National Institute of Standards and Technology (NIST). This research was further supported by NIST, the Defense Advanced Research Projects Agency SCOUT Program (W31P4Q-15-1-0011), and the Air Force Office of Scientific Research (FA9550-16-1-0016).


\bibliography{MyLibrary}

\end{document}